\documentclass[aip,apl,amsmath,amssymb,groupedaddress,print]{revtex4-1}

\usepackage{amsfonts}
\usepackage{amsmath}
\usepackage{amssymb}
\usepackage{graphicx, epstopdf, color}
\usepackage{bm}
 \usepackage[table,xcdraw]{xcolor}

\usepackage[normalem]{ulem}

\usepackage [colorlinks, linkcolor=blue, citecolor=blue]{hyperref}

\begin{document}

    \begin{center}
		\textbf{ \centerline{ \large Enhanced versatility of table-top X-rays from van der Waals structures
		}  } 
		
		Sunchao Huang$^1$, Ruihuan Duan$^2$, Nikhil Pramanik$^1$, Chris Boothroyd$^{3,4}$, Zheng Liu$^4$ and Liang Jie Wong$^{1\star}$
	\end{center}
	\textit{$^1$School of Electrical and Electronic Engineering, Nanyang Technological University, 50 Nanyang Avenue, Singapore 639798, Singapore \\
$^2$CINTRA CNRS/NTU/THALES, UMI 3288, Research Techno Plaza, Nanyang Technological University, 50 Nanyang Avenue, Singapore 637371, Singapore \\
	$^3$Facility for Analysis, Characterisation, Testing and Simulation (FACTS),
	Nanyang Technological University,
	50 Nanyang Avenue, Singapore 639798, Singapore \\
		$^4$School of Materials Science and Engineering,
		Nanyang Technological University,
		50 Nanyang Avenue, Singapore 639798, Singapore } \\
	$^{\star}$Email: liangjie.wong@ntu.edu.sg \\

Van der Waals (vdW) materials have attracted much interest for their myriad unique electronic, mechanical and thermal properties.
In particular, they are promising  candidates for monochromatic, table-top X-ray sources. This work reveals that the versatility of the table-top vdW X-ray source goes beyond what has been demonstrated so far. By introducing a tilt angle between the vdW structure and the incident electron beam, it is theoretically and experimentally shown that the accessible photon energy range is more than doubled.  This allows for greater versatility in real-time tuning of the vdW X-ray source. Furthermore, this work shows that the accessible photon energy range is maximized by simultaneously controlling both the electron energy and the vdW structure tilt. These results should pave the way for highly tunable, compact X-ray sources, with potential applications including hyperspectral X-ray fluoroscopy and X-ray quantum optics.

\section*{Introduction}

Van der Waals (vdW) materials are a distinctive family of materials consisting of two-dimensional sheets of atoms, either in single-layer form (e.g., graphene), or in multilayer form, held together by van der Waals forces (e.g., graphite). Members of this family exhibit unique properties that can include linear energy-momentum dispersion \cite{li2008dirac}, giant intrinsic charge mobility  \cite{geim2009graphene,zhao2017high}, extreme electromagnetic confinement \cite{jablan2009plasmonics, iranzo2018probing}, quantum Hall effects \cite{gusynin2005unconventional,
novoselov2007room, dean2013hofstadter}, van Hove singularities \cite{havener2014van}, in-plane and tunneling pressure sensors \cite{xu2011plane}, gate-tunable plasmons \cite{fei2012gate}, tunable photon polaritons \cite{dai2014tunable} and superconductivity \cite{yankowitz2019tuning}. 
Among its many exciting prospects, vdW materials are promising platforms for nanomaterial-based X-ray sources. For example, 
the nanoscale electromagnetic confinement achievable in 2D vdW materials like graphene makes them promising platforms for compact, free electron-driven sources of high-brightness X-rays \cite{wong2016towards,rosolen2018metasurface,pizzi2020graphene,rivera2019light}. Recently, the generation of tunable X-rays from free electron-driven vdW materials was theoretically predicted and experimentally
demonstrated \cite{shentcis2020tunable}.  
The output X-ray peaks can be tuned by controlling the electron kinetic energy and the atomic composition of the vdW material. X-ray generation via electron-crystal interaction also exists in conventional crystalline materials \cite{baryshevsky2005parametric,baryshevsky2007experimental,freudenberger1995parametric, brenzinger1997narrow,takabayashi2012observation, korotchenko2012quantum}. However, vdW materials and heterostructures \cite{geim2013van, jin2021heteroepitaxial,peter2019chiral,li2020general,
aubrey2021general,zhang2019graphene,liang2020van} are attractive platforms due to the large variety of compound combinations that provide control over the exact lattice constants determining the radiation spectrum. 
 Besides, vdW materials have no dangling bonds or reconstruction at the surface \cite{cho,liu}, and high in-plane thermal conductivities
 \cite{jiang2017probing}, making them a compelling basis for compact, versatile, high-quality X-ray sources \cite{shentcis2020tunable, balanov2021temporal}. \\

Here, we show that the versatility of the vdW-based free electron-driven X-ray source can be significantly enhanced by combining the aforementioned tuning mechanisms -- by electron energy and atomic composition -- with a third mechanism: by varying the tilt angle of the vdW structure, denoted  $\theta_{\text{til}}$ in Fig.~\ref{Fig1angle}a. Specifically, we theoretically predict and experimentally demonstrate that the range of accessible photon energies increases by over 100$\%$ when we simultaneously vary both the electron energy and the vdW structure tilt angle. This tilt angle is readily controlled by mechanically rotating the vdW structure with respect to the electron beam. In the process, we present a relativistic theory of free electron radiation in crystalline materials that accounts for all orders of emission processes within the same framework. We also demonstrate photon energy tuning via vdW structure tilt alone, showing that a wide range of photon energies can be accessed by varying the structure tilt angle at one fixed electron energy. Tuning the photon energy via the vdW structure tilt angle alone also has the advantage of being a simple mechanical maneuver that does not require re-stabilizing and realigning the electron beam, as is the case when the electron energy is varied. Our results should pave the way for greater versatility in compact X-ray sources based on vdW materials.

\section*{Results}
\begin {figure}[!hbt]
\begin {center}
\centering
\includegraphics [angle = 0, width = 1 \textwidth]{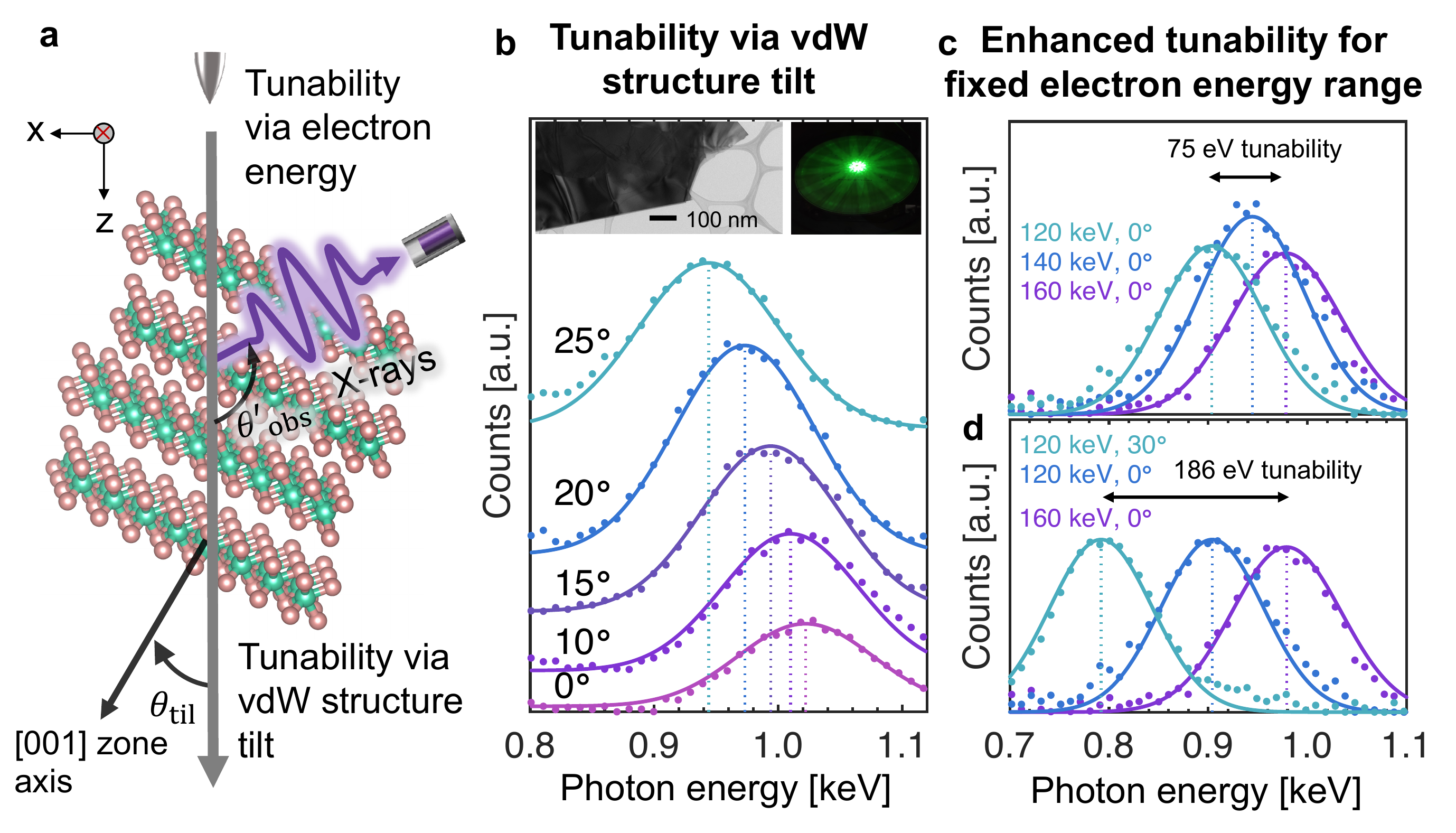}
\end{center}
\linespread{1}\selectfont{} 
\caption{ \textbf{ Enhanced tunability of X-ray emission from van der Waals (vdW) materials by simultaneously varying the vdW structure tilt angle and the electron energy. a,}  An incident electron beam scatters off the periodic lattice of a vdW material, generating X-rays via parametric X-ray radiation and coherent
Bremsstrahlung. \textbf{b} shows the spectra 
generated by a 200 keV electron beam impinging on a WSe$_2$ single crystal (left top insert shows its TEM image) at various tilt angles $\theta_{\text{til}}$,
defined as the angle between the incident electron beam and the [001] zone axis. The angle is calibrated to within 0.5 degrees of accuracy based on Kikuchi lines (right top
insert).  \textbf{c} and \textbf{d}, Enhanced tunability is achieved by simultaneously varying both the electron energy and the vdW structure tilt angle. The accessible
X-ray radiation photon energy range is more than doubled from 75 eV to 186 eV under the combined tuning scheme. 
In all panels, experimental results are represented by filled circles, and theoretical predictions by solid curves. Vertical dotted lines indicate the peak photon energy predicted by equation (\ref{EqPCBpeak}).
}
\label {Fig1angle}
\end{figure}

Figure~\ref{Fig1angle}a illustrates the vdW-based free electron-driven X-ray generation process.
The passage of a free electron through a multilayered vdW structure modulates the bound electrons of the material's atoms, creating polarization currents that emit radiation via parametric X-ray radiation (PXR). PXR can also be understood as the diffraction
of the incident electron’s Coulomb field off the periodic arrangement of atoms; in this respect, it is simply an atomic scale version of the Smith-Purcell radiation process \cite{112,yang2018maximal,ye2019deep}. At the same time, the free electron itself is modulated by the periodic potential of the atomic lattice, resulting in photon emission via coherent bremsstrahlung (CB). In particular, CB results from the interference of emitted radiation from multiple, periodically spaced Bremsstrahlung events. 
These two types of X-ray radiation (PXR and CB) share the same energy
peak at a given detection angle, and are collectively termed parametric coherent bremsstrahlung (PCB)
\cite{shentcis2020tunable}.  In Fig.~\ref{Fig1angle}a, the angle between the incident electron (along the z-direction) and the [001] zone axis is denoted $\theta_{\text{til}}$, and is henceforth referred to as the vdW structure tilt angle. \\

We perform the experiments in transmission electron microscopes (TEM), measuring the emitted X-rays using energy dispersive X-ray spectroscopy (EDS) detectors, as detailed in the Methods. We obtain the average radiation
intensity per electron of a large, incoherent electron beam as
\\
\begin{eqnarray}
	\bigg< \frac{\text{d}^2N}{\text{d}\omega \text{d}\Omega} \bigg> \approx \frac{1}{N_e} \frac{\alpha \omega}{4 \pi^2 c^2} \sum_{i=1}^{N_e} \bigg|
	\int_0^{t_L} \bm{v}_i(t) \cdot \bm{E}_{\bm{k}s} (\bm{r}_i,\omega) e^{-\text{i}\omega t} \text{d}t \bigg|^2 ,
	\label{Eqd2NdwdW}
	\label{eqIntensity}
\end{eqnarray}
\\
where $N$ is the number of emitted photons, $\omega$ is the angular
frequency of the emitted photon, $\Omega$ is the solid angle, $ N_e$ is the number of incident electrons, $\alpha$ is the fine-structure constant, $c$ is the speed of light in free space, $t_L$ is the interaction time of the electron with the crystal,
$\bm v_i(t)$ is the velocity of electron, obtained via the relativistic Newton-Lorentz equation, $\bm
E_{\bm{k}s}(\bm r_i, \omega)$ is an eigenmode of the crystal, $\bm k$ is the
wave vector of the radiation field, $s$ is the index of the polarization, and $\bm r_i$ is the trajectory of the electron.  Our derivation of equation~(\ref{eqIntensity}) is based on the scattering theory of Baryshevsky et al. \cite{feranchuk2000parametric,nitta1991kinematical,baryshevsky2005parametric}, but importantly goes beyond it by a) including relativistic corrections for the incident electron, b) summing over all the radiation arising from the various reciprocal lattice vectors $\bm g$, and c) averaging over the initial positions of the electron on the crystal surface. Although we focus on electrons in this study, our theory is valid for any charged particle when the corresponding values for charge and rest mass are used. Our approach has advantages over approaches that consider PXR and CB using separate theoretical frameworks\cite{shentcis2020tunable}, as we are able to capture the effects of interference between PXR and CB processes, as well as the presence of higher-order processes beyond PXR and CB. For details of the derivation, see Supplementary Section \uppercase\expandafter{\romannumeral1}.
We obtain the peak photon energy of the output X-rays from the result of equation~(\ref{eqIntensity}) as
\\
\begin{eqnarray}
	E = ~ \hbar c~ \frac{{\beta}_0 \hat{\textbf{z}}\cdot (\hat{U}\bm{g}_0 ) }{1-\beta_0 \cos\theta'_{\text{obs}}},  
\label{EqPCBpeak}
\end{eqnarray}
\\
where $\hbar$ is the reduced Planck constant, $ \beta_0 =  v_0/c$, $ v_0$ being the initial speed of the incident
electron, $\hat{\textbf{z}}\cdot (\hat{U} \bm{g}_0) = (-\sin\phi_{\text{til}}\cos\theta_{\text{til}})~g_{0x} + (\sin\phi_{\text{til}} \sin\theta_{\text{til}})~ g_{0y} + (\cos\theta_{\text{til}}) ~g_{0z} $, where $\hat{U}$ is the unitary matrix and $\bm{g}_0$ is the reciprocal lattice vector in the unrotated frame, i.e., when $\theta_{\text{til}} = \phi_{\text{til}} = 0^{\circ}$, $\phi_{\text{til}}$ is the rotation angle of the crystal with respect to the z-axis
 and $\theta'_{\text{obs}}$ is the effective angle between the electron beam and the observation direction as
 shown in Fig.~\ref{Fig1angle}a.   
Taking the X-ray peak broadening due to electron beam divergence, the detector energy resolution, and the shadowing effect into account, we obtain the following expression for the measured bandwidth of the PCB peaks (see Supplementary Section \uppercase\expandafter{\romannumeral1} for details):
\\
\begin{equation}
\begin{split}
	\Delta E_{\text{tot}} \approx \bigg[  \left( 5.6\frac{\hbar v_0}{L} \right)^2 +  \bigg( \frac{\hbar c \beta_0 \hat{\textbf{z}}\cdot 	\big({\partial \hat{U}\bm{g}_0}/{\partial \theta_{\text{til}}}\big) }{1-\beta_0 \cos\theta'_{\text{obs}}} \Delta \theta_{\text{e}} \bigg)^2   + R^2  +\\
	E^2\left(  \frac{ \beta_0 \sin\theta'_{\text{obs}} }{1-\beta_0 \cos\theta'_{\text{obs}}} \Delta \theta'_{\text{obs}} \right)^2    \bigg]^{1/2},
	\end{split}
\label{eqWidth}
\end{equation}
\\
where $L$ is the interaction length and $R$ is 
the energy resolution of the energy dispersive X-ray spectroscopy (EDS) detector. In determining the actual observation angle $\theta'_{\text{obs}}$ and its angular spread $\Delta\theta'_{\text{obs}}$ in equation~(\ref{eqWidth}), we take into account the shadowing effect, which causes the effective observation angle to increase (the effective observation angular spread to decrease) by a few degrees from its default value $\theta_{\text{obs}}$ ($\Delta \theta_{\text{obs}}$). This deviation is due to the edge of the sample holder partly blocking the output X-rays on their way towards the EDS detector  \cite{pantel2011coherent} (see Supplementary Section  \uppercase\expandafter{\romannumeral3} for more details). 
 The first term in equation~(\ref{eqWidth}) corresponds to the
intrinsic bandwidth of the PCB X-ray peak \cite{feranchuk2000parametric} obtained from equation~(\ref{eqIntensity}), which is on the order of 1 eV in our case. The second term corresponds to effects of electron beam divergence. In our experiments the beam divergence $\Delta \theta_{\text{e}} \approx 1$ mrad. The third term accounts for the energy resolution of the EDS detector.
 The final term accounts for the finite range of observation directions admitted by the
angular aperture of the EDS detector. 
 Figure~\ref{Fig1angle} shows good agreement between the experimental measurements (filled circles) with the predictions of our theory (solid lines). \\

\begin {figure}[!hbt]
\begin {center}
\centering
\includegraphics [angle = 0, width = 1.0 \textwidth]{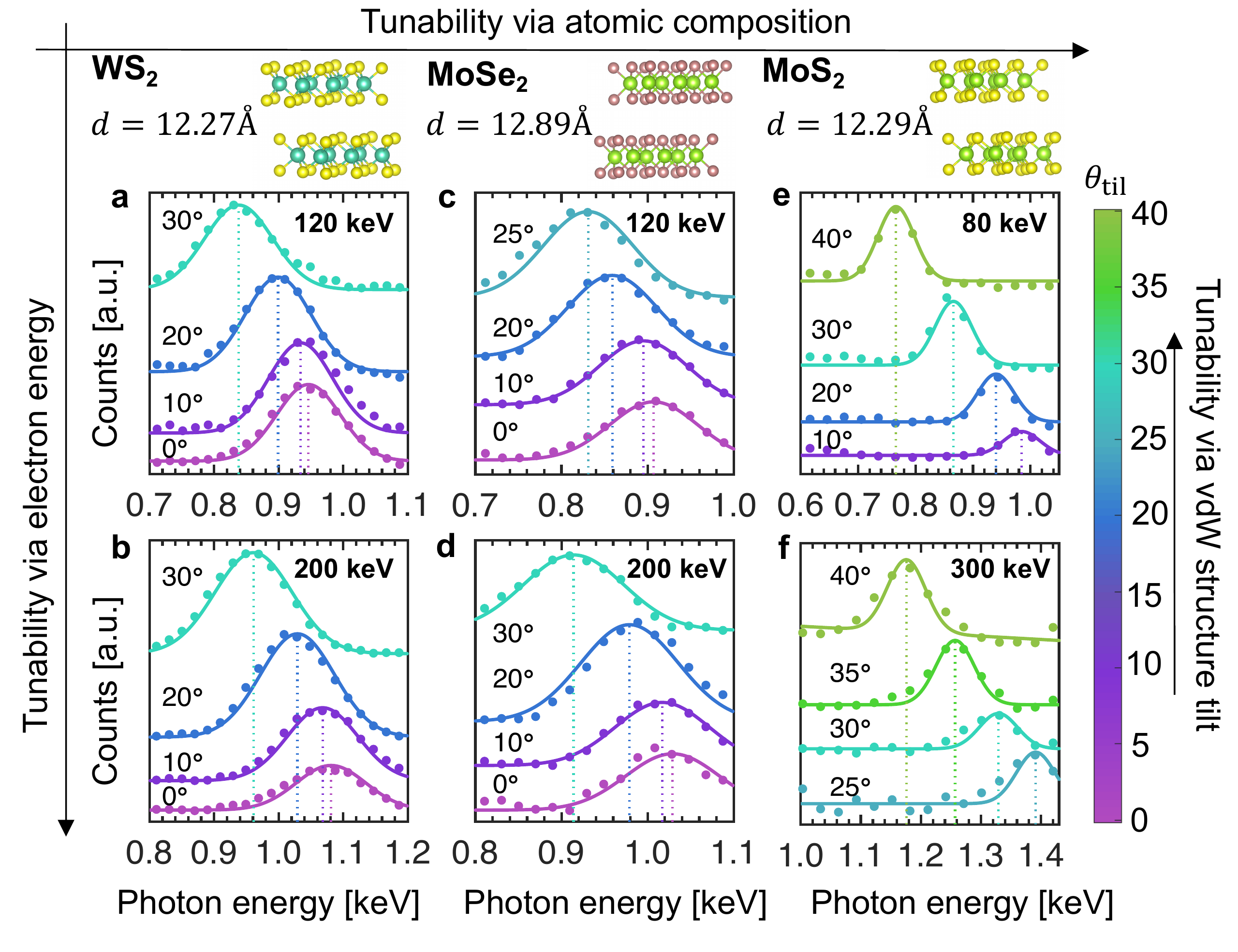}
\end{center}
\linespread{1}\selectfont{} 
\caption{ \textbf{Three-dimensional tunability of free electron
radiation in vdW materials: Tunability via vdW structure tilt, electron energy and atomic composition. a-f}  together illustrate our
paradigm of a highly versatile, compact X-ray source in which the photon energy can be tuned over a wide range by varying the structure tilt, the electron
energy and the atomic composition. 
X-ray spectra from WS$_2$ (\textbf{a,b}), MoSe$_2$ (\textbf{c,d}) and MoS$_2$(\textbf{e,f}) are tailored by varying electron energy (labelled at top right corner of each panel) and structure tilt angle $\theta_{\text{til}}$ (labels individual curves within each panel). Filled circles represent experimental measurements and solid curves corresponds to theoretical predictions. Vertical dotted lines indicate the peak photon energy predicted by equation~(\ref{EqPCBpeak}). Comparing (\textbf{b}) and (\textbf{d}),
 WS$_2$ with a smaller interlayer spacing generates harder X-rays compared to MoSe$_2$ under the same conditions, showing the tunability of vdW X-rays via
atomic composition. 
In (\textbf{a-d}), $\theta_{\text{obs}} \approx 112.5^{\circ}$ and $\Delta\theta_{\text{obs}} \approx 12^{\circ}$. In (\textbf{e,f}), $\theta_{\text{obs}} \approx 91.5^{\circ}$ and $\Delta\theta_{\text{obs}} \approx 1^{\circ}$. 
In all cases, the intrinsic bandwidth is about a few eV, but is broadened by the energy resolution of the respective EDS detectors.}
\label {Fig2angle}
\end{figure}

Figure~\ref{Fig1angle}b shows the PCB spectrum when
 a 200 keV electron beam is incident on a WSe$_2$ single crystal. The X-ray photon energy is tuned from 1026~eV to 946~eV when the tilt angle of the WSe$_2$ single crystal is varied from $\theta_{\text{til}} = 0^{\circ}$ to $\theta_{\text{til}} = 25^{\circ}$, where we determine $\theta_{\text{til}}$ to
an accuracy better than 0.5$^{\circ}$ in the experiments by using
Kikuchi lines (right insert in Fig.~\ref{Fig1angle}b) \cite{heilmann1982computerized, rauch2014automated}. This attests to the feasibility of dynamically tuning (i.e., tuning in real-time) the output photon energy via vdW structure tilt alone. This would be helpful in scenarios where other tuning mechanisms are not as readily available: for instance, tuning via the electron energy typically requires readjustment of the accelerating voltage and realigning of the electron beam; whereas tuning via atomic composition requires the growth of a completely new material. A TEM image of our WSe$_2$ sample is shown in the left insert of Fig.~\ref{Fig1angle}b. 
In Figs.~\ref{Fig1angle}c and d, we consider an electron energy range of 120 keV to 160 keV. Figure~\ref{Fig1angle}c shows that the achievable output photon energy range is 75 eV when $\theta_{\text{til}}=0^{\circ}$ and only the electron energy is allowed to vary.
In Fig.~\ref{Fig1angle}d, this range increases by over 100$\%$ to 186 eV when we allow the electron energy and the vdW structure tilt angle to simultaneously vary. \\

\begin {figure}[!hbt]
\begin {center}
\centering
\includegraphics [angle = 0, width = 0.95 \textwidth]{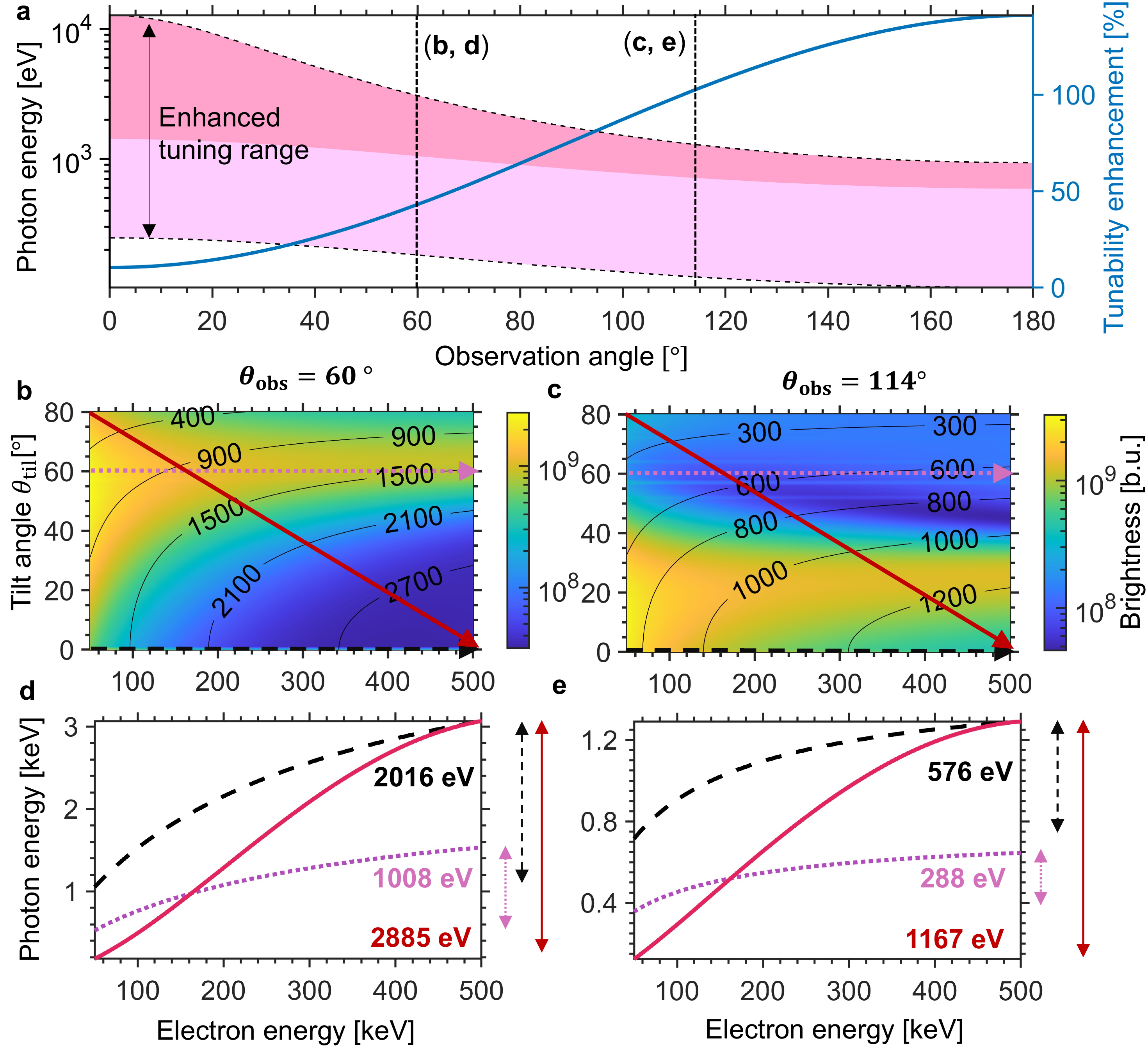}
\end{center}
\linespread{1}\selectfont{} 
\caption{ \textbf{Enhanced tunability for various observation angles in WS$_2$. a} 
The entire shaded region between the two dashed lines corresponds to the accessible photon energy range when vdW structure tilt and electron energy are simultaneously varied, whereas the darker pink shaded portion corresponds to that when only the electron energy is varied. The blue curve represents the percentage enhancement in the photon energy range in the former scheme. \textbf{b,c}  Brightness (equation~(\ref{eqIntensity})) as a function of electron energy for $\theta_{\text{til}}$ from $0^\circ$ to $80^\circ$ at different $\theta_{\text{obs}}$, where [b.u.] (``brightness units") stands for [photons $\text{s}^{-1}$ $\text{mm}^{-2}$  $\text{mrad}^{-2}$ per 0.1\% BW]. The photon energy of X-rays (in eV) is indicated by the black contour lines (equation~(\ref{EqPCBpeak})). \textbf{d,e} Accessible photon energy range by tuning along the arrows in colormaps \textbf{(b,c)} respectively. Our enhanced tunability especially favors emission at obtuse detector angles and in the soft X-ray range. In this figure, $\phi_{\text{til}} = 0^{\circ}$ and $L = 100 $ nm.
}
\label {PCB3}
\end{figure}

Dichalcogenide vdW materials like WSe$_2$, WS$_2$ and MoS$_2$ 
crystallize in a layered structure with slightly differing
interlayer distances \cite{chhowalla2013chemistry}, which offer opportunities to tune the output X-ray photon energy via atomic composition \cite{shentcis2020tunable}. Combined with tunability via the vdW structure tilt and the electron energy, this makes vdW materials a versatile platform for compact X-ray generation.
Figure~\ref{Fig2angle} shows three-dimensional tunability of the vdW X-ray
radiation: tunability via the electron energy,
tunability via the atomic composition, and tunability via the vdW structure tilt.
Tunability via the atomic composition allows the pre-customization of a PCB X-ray
source by choosing the constituents of the vdW structure. The many compound combinations possible in vdW materials provide precise control over the lattice constants that determine the radiation spectrum. On the other hand,
tunability via the electron energy and via the vdW structure tilt provide
dynamic tunability: the electron energy can be adjusted by changing the accelerator voltage of the electron source, and the vdW structure tilt angle can be adjusted by mechanical rotation.  It should be noted that the intrinsic bandwidth of the PCB peaks is also very narrow, being on the order of 1 eV in our regime of study. 
The measured bandwidth is significantly broadened 
by the large energy resolutions and observation angle spreads of the respective EDS detectors. \\

Figure~\ref{PCB3}a depicts the X-ray photon energy range accessible with the vdW-based X-ray source.
If only the electron energy is allowed to vary, only photon energies in the dark pink shaded region can be accessed. This region becomes increasingly narrow at larger observation angles, which favor softer X-rays that could be beneficial for biological imaging. On the other hand, if the electron energy is varied together with vdW structure tilt angle, we see that the accessible range of output X-ray photon energies expands to the entire pink-shaded region, bounded by the pair of dashed lines. The solid blue line in Fig.~\ref{PCB3}a reflects the percentage enhancement in accessible photon energy range by combining control over both electron energy and vdW structure tilt. This percentage enhancement is well in excess of $100\%$ at larger detector angles.
Here, we have considered an electron source that can be tuned from 50 keV to 500 keV (which covers the electron energy range of most TEMs). Figures~\ref{PCB3}b,d and \ref{PCB3}c,e focus on the specific cases where $\theta_{\text{obs}} = 60^{\circ}$ and $\theta_{\text{obs}} =  114^{\circ}$ respectively. In both cases (as in all other cases used in  Fig.~\ref{PCB3}a), the tuning scheme via electron energy and vdW structure tilt runs diagonally across the range of vdW structure tilt and electron energies considered (red lines in Figs.~\ref{PCB3}b,c). The resulting photon energy peaks are shown in Figs.~\ref{PCB3}d,e  respectively, and contrasted against cases where only the electron energy is allowed to vary (horizontal lines in Figs.~\ref{PCB3}b,c). At the same time, the colormaps in Figs.~\ref{PCB3}b,c show the brightness of the output X-ray photons, as calculated from equation~(\ref{eqIntensity}). We see that the brightness can vary significantly across the entire tuning range. For any specific output X-ray photon energy, it is possible to maximize the X-ray brightness with the freedom to vary both electron energy and vdW structure tilt angle. Simultaneously controlling both
electron energy and vdW structure tilt thus allows us to optimize the accessible photon energy range as well as the intensity of the vdW X-ray source. \\

For relativistic electrons (1-10 MeV), tuning by varying the electron energy becomes challenging  at observation angles beyond $20^{\circ}$, as discussed in Supplementary Section \uppercase\expandafter{\romannumeral2}. The only feasible way to tune the photon energy in real-time for relativistic electrons is via the vdW structure tilt angle. Specifically, tuning via the vdW structure tilt angle allows us to enhance the emitted photon energy range by 1873\% and  654\% for $\theta_{\text{obs}} = 114^{\circ}$ and $\theta_{\text{obs}} = 60^{\circ} $ respectively, compared to tuning by varying the electron energy.

\section*{Discussion}
The vdW X-ray generation scheme we study is highly complementary to other existing methods of X-ray generation. The vdW X-ray source is dynamically tunable in frequency, unlike
traditional X-ray tubes whose output peaks are fixed at the characteristic
frequencies of the anode material \cite{schweppe1994accurate}. Furthermore, it
requires neither highly relativistic electrons nor high intensity lasers, as
in undulator-based X-ray sources \cite{bassler1999soft} and high-harmonic
generation \cite{rosolen2018metasurface}. 
Our results should pave the way for the realization of dynamically tunable, compact X-ray sources, which have a wide range of potential applications in imaging and inspection \cite{carroll2002tunable, hara2013two, hayakawa2013x}, including X-ray hyperspectral imaging and X-ray quantum optics \cite{wong2021p,nordgren1995soft,jannis2019spectroscopic,sofer2019quantum}. 
The study of shaping of incident free electrons \cite{mihalcea2014beam,wong2021control,lemery2015tailored,
zhao2021quantum,karnieli2021superradiance,kfir2021optical,
di2020free,karnieli2021coherence,graves2012intense,
luiten2004realize,engelen2013high} -- on the level of either the macroscopic bunch structure or the individual electron wavefunction -- is a subject of active investigation that could lead to greater control and enhancement of the output radiation. \\

In our experiments, we measured 7.9$\times 10^{4}$ PCB photons over a duration
of 1000 s (live time) from WS$_2$ at $\theta_{\text{til}} = 30^{\circ}$, shown
in Fig.~\ref{Fig2angle}b. This yields a flux of 79 photons s$^{-1}$, which is
in excellent agreement with our theory for a current of 0.34 nA. The
relatively low electron current was used to avoid pileup effects during the
measurement of X-rays by the EDS detector, whose dead time was kept below 30\%.
This scales to a brightness of $\sim$  1$\times10^9$ photons s$^{-1}$
mm$^{-2}$ mrad$^{-2}$ per 0.1\% BW when an electron beam of 1 nA current and 1
nm spot size is employed, which is consistent with the results reported in
Ref. \cite{shentcis2020tunable}. This brightness also compares favorably with
that of high harmonic generation ($10^{5}-10^{12}$ photons  s$^{-1}$ mm$^{-2}$
mrad$^{-2}$ per 0.1\% BW) in the water-window
\cite{helk2019perspective,kordel2020laboratory}. The angular flux density from
our unoptimized source is about $\sim$ 4$\times10^6$ \textit{I}(A) [photons
s$^{-1}$ mrad$^{-2}$ per 0.1\% BW], which already comes close to that of
conventional X-ray tubes $\sim$ $10^7-10^8$ \textit{I}(A) [photons s$^{-1}$
mrad$^{-2}$ per 0.1\% BW] \cite{feranchuk1999new}, where \textit{I}(A) refers
to the current in Amperes.   
To enhance the performance of our X-ray source further, it should be noted that the peak brightness is directly proportional to the electron current, and to the square of the interaction length (i.e., $\propto$ \textit{L}$^2$). As in X-ray tubes, a larger electron current will generate more heat and increase the possibility of thermal damage. In this regard, van der Waals materials like graphite have an advantage over conventional materials (e.g., tungsten, commonly used as the anode in X-ray tubes) due to the former's superior thermal conductivity and melting point. Increasing the thickness of the target material increases the interaction length \textit{L}. Innovative methods to increase the interaction length include having the electrons travel near the edge of vdW materials in a Smith-Purcell-like configuration that has been termed edge PXR \cite{balanov2021temporal}. This allows the electron's Coulomb fields to scatter off the crystal lattice while minimizing collisions of the electrons themselves with the material. \\

In conclusion, we have shown that the versatility of the vdW-based free electron X-ray source can be  significantly enhanced with the introduction of a new control parameter: the vdW structure tilt angle, which can be varied in real-time by mechanically rotating the vdW target with respect to the electron beam. Specifically, we show that the range of accessible photon energies increases by over $100\%$ when we simultaneously vary both the electron energy and the vdW tilt angle. At the same time, we present a relativistic theory of PCB that not only accounts for both PXR and CBS in the same framework, but also includes arbitrarily higher-order free electron radiation processes. This, combined with the ability to tailor the vdW-based X-rays via atomic composition, makes van der Waals materials a promising platform for highly versatile, tunable X-ray sources. Our results also show that a wide range of photon energies can be accessed just by varying the vdW tilt angle alone, even with a fixed electron energy and atomic composition. Although our study focuses on moderate electron energies (0.05-10 MeV), our method of enhancing the
photon energy range by combining control over electron energy and tilt angle applies to other ranges of electron energies, and also other crystalline material systems
beyond vdW materials. Our results should pave the way to realizing compact sources of high quality X-rays for applications including hyperspectral X-ray fluoroscopy and X-ray quantum optics.

\section*{Materials and Methods}

\textbf{Sample preparation:}
2D bulk MX$_2$ (M= Mo, W; X = S, Se) single crystals were synthesized by the 
normal chemical vapor transport method. The stoichiometric ratio of high purity M and X with a bit of iodide as transport agent are loaded in a silica tube, which is sealed in a high vacuum environment. The sealed silica tube is loaded in a two-zone furnace, whose growth zone is heated to 850~$^\circ$C and reaction zone is heated to 950~$^\circ$C within 24 hours, and held for ten days.  Finally, bulk MX$_2$ single crystals are collected in the growth zone. The few-layer MX$_2$ nanoflakes are exfoliated mechanically onto silicon substrates (covered with a 285 nm SiO$_2$ film), and transferred to Au grids (for TEM measurement) with the aid of the wet-transfer method. \\

\textbf{X-ray measurements:} The vdW-based X-ray emission measurements were conducted in transmission electron microscopes (TEM). A highly collimated electron beam is sent towards the vdW material in the sample holder, which can be tilted. The emitted X-ray spectra were measured using a silicon drift energy dispersive X-ray spectroscopy (EDS) detector. The EDS detector was calibrated by ourselves to enable measurement of X-ray peak energies with an accuracy of $\pm$2.5 eV (see Supplementary Section \uppercase\expandafter{\romannumeral4} for details).  
The experiments shown in Fig.~\ref{Fig1angle} and
Figs.~\ref{Fig2angle}(\text{a-d}) were conducted in a JEOL 2010HR TEM, which uses 120--200 keV electrons.
In our photon energy range of interest (0.7 keV-1.4 keV), the energy resolution is $R \approx 97$ eV for the EDS detector in the JEOL 2010HR TEM. The detector's observation angle and observation angle range are $\theta_{\text{obs}} \approx 112.5^\circ$ and $\Delta\theta_{\text{obs}} \approx 12^{\circ}$ respectively. The experiments in Figs.~\ref{Fig2angle}(\text{e-f}) were performed in a JEM-ARM300F TEM, which uses 80 keV and 300 keV electrons. In our photon energy range of interest, energy resolution $R \approx 75$ eV for the EDS detector in a JEM-ARM300F TEM. The background radiation of the measured spectra was subtracted using NIST DTSA-II\cite{ritchie2012eds, newbury2015performing}. In both TEMs, the sample holder is made of beryllium, and can be rotated about the x- and y-axes (the x-y plane being that which lies parallel to the surface of the sample holder), allowing us to determine $\theta_{\text{til}}$ to an accuracy better than 0.5 degrees with the help of Kikuchi lines. In all measurements, increasing $\theta_{\text{til}}$ further tilts the sample towards the EDS detector. The range of $\theta_{\text{til}}$ is $\pm 30^{\circ}$ and $\pm 40^{\circ}$ for the JEOL 2010 HR TEM and the JEM-ARM300F TEM, respectively. \\

\textbf{Theory and simulations:}
See Supplementary Section \uppercase\expandafter{\romannumeral1} for details.

\section*{Acknowledgements}
We would like to Yee Yan Tay, Michael Shentcis and Ido Kaminer for helpful discussions. We would like to acknowledge the Facility for Analysis, Characterisation, Testing and Simulation, Nanyang Technological University, Singapore, for use of their electron microscopy and X-ray facilities. 
\textbf{Funding:} This project was supported by the Agency for Science, Technology
 and Research (A$*$STAR) Science $\&$ Engineering Research Council (Grant No. A1984c0043).
LJW acknowledges the Nanyang Assistant Professorship Start-up Grant.
\textbf{Author contributions:}
SH led the project and analysed the data. SH designed and performed the X-ray measurements with the help and advice of CB, who also performed some of the measurements. RD prepared all the samples with the help of ZL. NP  developed the theory and performed the simulations with the help of SH and LJW.
SH, NP and LJW wrote the paper, with inputs from all other authors.
LJW conceived the idea and supervised the project. 
\textbf{Competing interests:}
The authors declare no competing interests. 
\textbf{Data availability:}
All data needed to evaluate the conclusions in the paper are present in the paper and/or the Supplementary Materials.
Additional data that support the plots within this paper are available from the corresponding authors upon reasonable request.

\bibliographystyle{unsrt}

\end{document}